\documentclass{article}
\usepackage{graphicx}
\usepackage{subfig}
\usepackage{authblk}

\title{Simplify Node-RED For End User Development in SeismoCloud}

\author[1]{Enrico Bassetti}
\author[1]{Edoardo Ottavianelli}
\author[1]{Emanuele Panizzi}
\affil[1]{Computer Science department, Sapienza University of Rome,
  Piazzale Aldo Moro 5, 00185 Roma, Italy}
\date{}

\begin{document}

\maketitle

\begin{abstract}
Networks of IoT devices often require configuration and definition of behavior by the final user. Node-RED is a flow-based programming platform commonly used for End User Development, but it requires networking and protocols skills in order to be efficiently used. We add a level of abstraction to Node-RED nodes in order to allow non-skilled users to configure and control networks of IoT devices and online services. We applied such abstractions to the SeismoCloud application for earthquake monitoring.
\end{abstract}

\section{Introduction}
Smart objects and Internet of Things (IoT) devices are becoming widespread, thus they more and more often require that end users define their behavior and configure them to connect with other objects as well as with online services.

Programming tools exist that allow wiring together hardware devices, APIs and online services, offering a powerful mechanism that provides full control to the end user, but that often require users to be knowledgeable about technical issues.

Node-RED \cite{nodered} is a flow-based programming \cite{morrison2010flow} platform, commonly used for End User Development. It is built as an event-driven platform which routes messages between \textit{nodes}. A group of nodes that execute one or more actions starting from an event is called \textit{action flow}. These \textit{nodes} are provided officially by Node-RED team, or they can be added as third-party plugins. This makes Node-RED highly extendable, while remaining very simple as approach.

However, Node-RED's ``End User''  is a person who has some knowledge about underlying protocols such as MQTT, HTTP requests, WebSockets, etc. In fact, works in literature that use Node-RED are requiring some prior knowledge of underlying protocols (for example, in Tabaa et al. \cite{tabaa2018industrial}, skills on Modbus standards and protocols are required).

The purpose of our contribution is to allow non-technical users to use Node-RED to configure and control networks of IoT devices and online services. This has been shown to be viable using abstraction with domain specific items\cite{ghiani2017personalization}.

To reach our goal, we adapted Node-RED to provide the end user with nodes that have a higher level of abstraction, so that non-technical users can configure and personalize their IoT ecosystem, still maintaining deep control over it.

\section{Related works}

Node-RED is widely used as EUD interface for different projects, ranging from simple IoT sensors integration (like temperature sensors \cite{lekic2018iot} or air quality monitors \cite{chanthakit2018mqtt}) to home automation \cite{rajalakshmi2017internet}, to complex industrial automation \cite{tabaa2018industrial}. It is also used to teach in Data Engineering courses \cite{chaczko2017learning} due to its simplicity.

A similar product is \textit{IFTTT} ``\textit{if this then that}''\cite{ovadia2014automate}, which is a cloud-based \textit{trigger-action} End User Development platform. However, \textit{IFTTT} supports only simple flows\cite{ur2014practical} (from triggers to action), and not complex flows as Node-RED does (for example, it is impossible to synthesize in IFTTT an industrial automation as in Tabaa et al. \cite{tabaa2018industrial}).

\textit{Ghiani et al.}\cite{ghiani2017personalization} provides an extensive and detailed analysis on the mechanisms behind EUD when targeting an end-user. Node-RED itself already implemented some of these techniques. However, while \textit{Ghiani et al.} ``TARE'' system is using a composition technique with buttons to build rules, our proposal uses a visual approach, based on the idea of "information flow".

\section{Application scenario}

The project where we applied Node-RED is SeismoCloud \cite{panizzi2016seismocloud}, a low-cost earthquake early warning system, based on smartphones and \textit{Internet-of-Things} devices. SeismoCloud generates events from different sources, such as official earthquake feeds, vibrations detected by sensors deployed by users in a crowd-sensing fashion, and metadata (sensor temperature, status, etc). It uses these sources internally to store and render data, and to generate earthquake early warnings. 

Node-RED is the core of the new End User Development platform in SeismoCloud, allowing users to create custom flows\footnote{Node-RED user can drag and drop nodes into one or more flows (group of nodes for specific purposes, e.g. room light control), and link the output of a node to the input of another node allowing a message ``produced'' by a node to be ``consumed'' by the linked one (which, in turn, can output another message).} in order to access and use the data provided by the different sources above and create personal online services (e.g. alarms, statistics, etc.). As our users do not generally have computer science knowledge, our goal is to provide a tool that everyone can use, without requiring technical skills. However, to reach this level of abstraction, we had to modify Node-RED's ``tool set'', removing any needs of knowledgeability and understanding of IoT- and network-related concepts.

\section{Adding abstractions in Node-RED}

A SeismoCloud sensor, just like the majority of IoT sensors, uses MQTT for signaling/data. Even if MQTT is a very simple protocol, all specific configuration about the MQTT connection, topic subscription and publishing may be out of range for a normal user. Also, some information about earthquakes are available via REST API calls (for mobile apps and website, mostly). These technologies require at least some programming skills to be used.

\begin{figure}
    \centering
    \subfloat[Temperature output node]{
        \includegraphics[width=3cm]{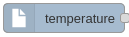}
        \label{fig:temperature}
    }
    \hspace{1cm}
    \subfloat[Perceptible earthquakes (per sensor)]{
        \includegraphics[width=3cm]{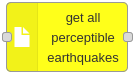}
        \label{fig:perceptible}
    }
    \caption{Example of two new nodes created for this experiment}
\end{figure}

In order to solve this problem, we created some domain-specific \textit{nodes} that users can drag and drop and configure just specifying domain-specific information in human-readable form. These nodes are abstractions for the Node-RED supported technologies, like MQTT or REST. For example, the ``temperature'' node (fig. \ref{fig:temperature}) can be easily configured just specifying the IoT sensor name, without any knowledge of the MQTT configuration which is automatically performed.

In fact, when the IoT sensor sends its temperature using MQTT to all its subscribers, this value is received by all the ``temperature'' \textit{nodes} (fig. \ref{fig:temperature}) configured using its name. All details such as \textit{MQTT broker}, credentials, topic name, QoS, TLS configuration, etc., are hidden from the point of view of the user.

This simplifies the interaction with Node-RED and allows anyone to use its EUD interface.

\begin{figure}
    \centering
    \includegraphics[width=12cm]{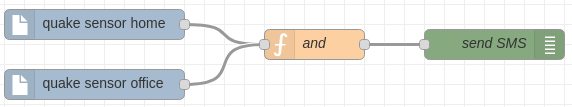}
    \caption{A very simple flow in Node-RED. Messages flows from left to right.}
    \label{fig:basic_flow}
\end{figure}

By mixing Node-RED standard nodes and our high level abstraction non-skilled people can build and understand Node-RED flows, such as ``when two devices emits a vibration message, send an SMS'' (see figure \ref{fig:basic_flow}).

\section{User testing and final results}

To evaluate whether our approach represents a sufficient simplification for standard users and an enabler for the EUD in SeismoCloud, while maintaining enough expressive power to create flows as desired by users, we planned and executed a test with 7 non-skilled users plus 1 expert user.


Users were asked to execute 4 tasks with different level of difficulty which required to get a data source event (e.g. a seismometer vibration detection or a temperature sensor above-threshold) and trigger a message using a simple label-based template. They were expected to drag and drop the proper nodes from the palette to the workspace and create a Node-RED flow.

For all trials we recorded the duration, success or failure in achieving task goals, whether the user understood newly introduced nodes. Also, users were asked to think aloud.

After performing tests with the first 4 users we made a second version of the nodes, simplifying the labels that describe the information required for their configuration.
This problem was found asking the users if they were understanding what they were doing. All the non-skilled users replied they understood that node returned a value, but the meaning wasn't clear.

In the following tests, all the users were able to complete the tasks without significant problems, although sometimes it took more time than we expected.

The result of this test is encouraging.
We are currently planning a test with expert users from our network of beta testers, in order to asses if the current interface is expressive enough and to what extent the underlying Node-RED built-in nodes are still necessary for operation in the SeismoCloud environment.

\bibliographystyle{abbrv}
\bibliography{bibliography}

\end{document}